\documentclass[preprint,12pt]{elsarticle} 
\usepackage{siunitx}
\usepackage{graphicx} 
\usepackage{subfig} 
\usepackage{pythontex}
\usepackage{amsmath}
\usepackage{mathrsfs}
\usepackage{pgfplots} 
\usepackage{pgfplotstable}  
\pgfplotsset{compat=1.3} 
\pgfplotsset{every axis legend/.append style={
    at={(1.05,1)},
    anchor=north west,font=\small}}
  
\usepackage{amssymb}
\usepackage{lineno}
\usepackage[absolute]{textpos}
\begin{document}

\begin{frontmatter}
\title{Numerical modeling of microstructure evolution in nanocrystalline alloys - grain boundary segregation, solute drag and mechanics}
\author{Prakarsh Pandey and Shiva Rudraraju} 
\address{Department of Mechanical Engineering, University of Wisconsin-Madison, Madison, WI, USA}

\begin{abstract}
Nanocrystalline (NC) alloys hold significant promise as structural alloys due to their superior mechanical properties over the traditional coarser grained microcrystalline alloys. Strength of metallic alloys is related to the underlying microstructural grain size - as represented by the classical Hall-Petch relation. Generally, a metals strength increases with decreasing mean grain size from the micrometer scale to the nanometer scale, until about a mean size of 20 nm. Any further decrease of grain size results in decreasing strength. Thus, there is an optimal range of mean grain size for most metals (about 20nm-10nm) about which maximum material strength can be obtained. In the context of NC alloys, stabilization of the grain size in this optimal range is one of the primary synthesis challenges. A large volume fraction of NC alloy microstructure is occupied by GBs. Since GBs increase the internal energy of the system through GB surface energy, during solidification and grain growth phase there is tendency to minimize GBs through grain coarsening. However, since smaller nanometer scale mean grain sizes are desired, certain phenomena like GB solute segregation and solute precipitation are utilized during alloy synthesis to mitigate grain growth below an optimal size. Numerical modeling these phenomena of GB-solute interactions and the evolution of these stabilized GBs under mechanical load are of immense interest to the NC alloy community. To enrich the numerical modeling formulations available in this space, we present here a phase-field method based numerical framework to model GB segregation, solute precipitation and effect of external loading on NC alloys. While some of these effects have been modeled in isolation, a unified treatment of the solute-GB segregation-mechanics interactions has not be considered in the literature. We present a three dimensional, FEM method based, finite-strain phase-field formulation for modeling grain evolution and microstructure stabilization in NC alloys.  Beyond the formulation and its computational implementation, various case studies demonstrate the applicability of this framework. Further, thermodynamic and kinetic arguments are provided based on the evolution of GB energy to explain the effects of solute drag, GB pinning and mechanical deformation. 
\end{abstract}


\end{frontmatter}
   
\section{Introduction}
\label{S:1} 

\par
Mechanical properties manifested by metallic alloys are a result of the load bearing ability of the underlying microstructure. The deformational response and related morphology evolution (grain evolution, dislocations, voids/cracks, phase transformations, etc.) at the grain-scale are the primary mechanisms that contribute to the structural scale elasto-plastic, fracture, fatigue and creep responses. Therefore, microstructure evolution at the grain-scale has long been studied in order to understand and predict the macroscopic mechanical properties. It is well known that optimal size, orientation and distribution of grains can improve material properties to a significant degree - resulting in a constant search for optimal synthesis, heat treatment and deformational processes modulating grain size through alloying, grain growth, refinement and recrystallization. Thus, the study of optimality and control over grain morphology attracts constant interest from the material synthesis and manufacturing community. Increasing the GB density is understood to yield a significant improvement in mechanical properties (elasto-plastic response; fracture and fatigue response; wear and tribological behavior) of alloys ~\cite{La2005, Szlufarska2005, Meyers2006, Hirth1972}. Examples of materials with high GB density include multilayered systems and nanocrystalline materials with layer thickness or grain sizes of the order of tens to hundreds of nanometers, wherein the increased grain interfaces block or impede dislocation motion leading to changes in the plastic strain accommodation in the material. 
\par

Controlled grain growth during material synthesis can be achieved through mitigating GB migration velocity or changing the orientation of GB migration. Both of these effects can be achieved through an enhanced understanding and control over GB interactions with solute atoms dissolved in the alloy.
It has long been known that interaction between solute and GBs can increase the recrystallization time ~\cite{LuckeAndMasing1956}. This phenomenon became known as solute drag, and well known mechanisms like Zener pinning~\cite{hillert1988-GG-inhibition, Smith1948, Hassold1990particleSizeEffect} and solute segregation~\cite{Kirchheim2002segregation, Kirchheim2007segregation} are known to \textcolor{black}{affect GB migration. For easier understanding of the interaction between solute and grain boundaries, it is convenient to classify solute particles into two categories i) stationary solute particles and ii) moving solute particles. Both of these categories have been intensively studied and interaction of stationary solute particles with GB is known as Zener pinning, whereas interaction of moving solute particles is better known as solute drag effect. In this paper we focus on the interaction of moving solute atoms with GBs.}
First theoretical treatment for solute drag was presented by ~\citet{luckeAndDetert1957} who showed that drag force was caused due to diffusion of solute atoms into the GBs. A rigorous thermodynamic analysis of this phenomenon was presented by ~\citet{Cahn1962ImpurityDrag}, showing that composition profile is symmetric for a stationary GB and drag force is generated due to asymmetry generated in the composition profile due to moving GB, and GB migration velocity can be a nonlinear function of driving force due to segregation, which under isotropic assumptions is linear.
Cahn also predicted that GB migration kinetics can exist in two regimes a) low velocity regime, in which GB migration is governed by diffusion of solute atoms into GBs, and b) high velocity regime, in which diffusion is not necessary and GB migration is governed by atomic jump across the GB, and effect of GB velocity and solute diffusivity on drag force are opposite to each other. 
Hillert and Sundman ~\cite{hillertAndSundman} later proposed their solution for steady state composition profile for a regular solution based on Gibb's free energy dissipation rather than Cahn, whose study was based on drag force. Hillert's method is equivalent to Cahn for a dilute solution with an advantage that it can be applied to phase transformations using Gibb's free energy dissipation.
\par

Various numerical studies have been performed to study the behavior of solute drag on solidification~\cite{wheeler1993drag-solidification, ahmed1998soluteDragOnSolidification}, solid state phase transformations~\cite{Loginova2003}, and on moving GBs. Increasingly, diffuse interface methods like phase-field modeling are the numerical method of choice for modeling microstructure evolution phenomena.  To model solute drag effect in a phase-field framework, segregation of solute atoms to the GB and diffusion within the GB must be considered. Cha \cite{Cha2002PF-on-MGB} proposed the first phase-field model that included solute drag effect. Later,  
Strandlund\cite{Strandlund2008EffectiveMobility} presented an effective mobility approach using the model by Odqvist\cite{Odqvist2003Eq.AtPhaseInter}. They suggested that drag effect can be captured in a phase-field representation if physically realistic mobilities are considered.
However, in these models solute drag effect was not explicitly considered. Gr\"{o}nhagen and \.{A}gren \cite{GronhagenAgren2007soluteSegregation} were successful in modelling solute drag effect by introducing a solute-composition dependent double-well potential in the Gibb's free energy expression and using diffusion flux for solute given by  Onsager's linear law of irreversible thermodynamics. They showed that their ``dynamic solute drag'' model was consistent with Cahn's and could describe dynamics of solute segregation to a stationary boundary as well as solute drag on a moving GB. Kim and Park \cite{Kim&park2008abnormalGG} demonstrated that model presented by Gr\"{o}nhagen and \.{A}gren could quantitatively describe solute drag and GB migration. They extended the model further to study abnormal grain growth, successfully showing that abnormal grain growth could be described by solute drag effect similar to particle pinning which has been traditionally used to study abnormal grain growth. Li et.al. \cite{LiWang&yang2009} also used the same model and were able to show the solute composition at the moving GB could increase with increased GB migration velocity and can be larger than equilibrium value, which has been experimentally observed \cite{Kasen1972highSegregation, song&Su1989nonEqModel, Jahazi2002nonEq}, but could not be explained through Cahn's theory.

\par

 These models studied solute segregation and solute acting on the GB in a dilute solution in which the solute was assumed to be completely miscible at all compositions. It has been found that solute segregation is the dominant phenomenon leading to reduction in GB energy in alloys with compositions away from the dilute limit \textcolor{black}{ and in literature the term 'solute drag' has been primarily used to explain this behavior}. However, there have been experimental evidences for nano-crystalline alloys in which solute is not miscible and system phase separates into solute rich and solute depleted zones through spinodal decomposition in the alloys with increased solute compositions. In immiscible alloys, both GB solute segregation and particle pinning due to precipitated solute rich regions affect grain growth. Most analytical and simulation studies have assumed immobile particles for Zener pinning, and while studying GB segregation - they do not consider immiscible alloys and precipitated solute particles. Effect of precipitated solute rich regimes to explain the stability of nano-crystalline alloys was studied by Murdoch and Schuh \cite{murdoch2013stabilityNC-phaseseparate}, who developed stability maps for nano-crystalline alloys explained by GB heat of segregation and alloy heat of mixing, and divided these maps into three regimes: a) where GB segregation does not result in stable structure, b) where macroscopic phase-separation is preferred despite the stability against grain growth, and c) where system is stable against grain growth and phase separation.

An effective phase-field model incorporating solute segregation and precipitation was presented by Fadi et. al. \cite{fadi2017immiscibleNC}, who studied the evolution of grain microstructures in terms of heat of segregation and heat of mixing - resulting in phase separation of solute atoms between the grains and the GBs. They consider a composition dependent double-well potential in the Gibb's free energy expression, and the solute evolution was governed by the conserved Cahn-Hilliard \cite{cahnHilliard1958} equation. We use this approach as a motivation for our current work.  
\par

The driving force for GB migration can be derived from the GB surface energy, lattice defect energy and stored elastic energy. Applying external deformation on a microstructure can cause significant changes in GB migration as shown by Sch\"{o}nfelder \cite{schonfelder1997MDforGG}, who uses molecular dynamics simulations of a copper bi-crystal to demonstrate that a driving force can be generated by applying anisotropic strain, provided energy densities are different in both grains.
However, the rates at which GB surface energy, defect energy and elastic energy can be minimized dictates their relative contribution to the driving force. If stored elastic energy can be minimized further than the GB surface energy, then grain growth can be considered to be significantly driven by the stored elastic energy.  A polycrystalline phase-field model was presented by Tonks et. al. \cite{tonks2011polycrystalline} who demonstrated that the amount of applied deformation accelerates the grain growth, compared to a relaxed polycrystal, in low temperature regions. However, in high temperature regions, grain growth is accelerated but the effect of applied deformation is reduced.
\par 
To enrich the numerical modeling frameworks available in this space of modeling microstructure evolution in NC alloys, we present here a phase-field method based numerical framework to model GB segregation, solute precipitation and effect of external loading on NC alloys. While some of these effects have been modeled in isolation, a unified treatment of the solute-GB segregation-mechanics interactions has not be considered in the literature. Further, the existing treatments are based on infinitesimal-strain formulation of mechanics and mostly developed in the finite difference methods (FDM) framework. This work derives from earlier advances on the phase-field modeling front presented by Gr\"{o}nhagen and \.{A}gren \cite{GronhagenAgren2007soluteSegregation}, Fadi et. al. \cite{fadi2017immiscibleNC} and Tonks et. al.\cite{tonksMillet2010}, and seeks to present a three dimensional, FEM method based, finite-strain phase-field formulation for modeling grain evolution and microstructure stabilization in NC alloys. Beyond the formulation and its computational implementation, various case studies demonstrate the applicability of this framework. Further, thermodynamic and kinetic arguments are provided based on the evolution of GB energy to explain the effects of solute drag, GB pinning and mechanical deformation.

\section{Free energy based representation of nanocrystalline grain evolution}

In this work, we adopt a phase-field based numerical treatment for modeling grain evolution, and the related phenomena of GB segregation and solute drag that are relevant to nanocrystalline alloys. The phase-field method is a popular free energy based treatment of microstructure evolution and the underlying interface kinetics. It has been widely used for modeling metallic solidification~\cite{kobayashi1993modeling, boettinger2002phase, bhagat2022modeling}, solid-state phase transformations~\cite{nishimori1990pattern, wang1998field, jiang2016multiphysics}, grain growth~\cite{chen1994computer, dewitt2020prisms}, etc., to name a few related applications. In the phase-field method, grain scale microstructure is numerically represented with a set of conserved or non-conserved (structural) order parameters that change continuously in space across diffuse GBs. The interfacial kinetics is modeled by evolving the phase-field variables via parabolic partial differential equations of diffusion - either the classical Fickian kinetics, or the more involved Cahn-Hilliard kinetics~\cite{cahnHilliard1958} (for conserved order parameters) and Allen-Cahn kinetics~\cite{allenCahnEquation} (for non-conserved order parameters).

Using a phase-field approach, we model multigrain microstructure evolution by associating each grain with an order parameter that takes on the value of unity within the grain, and a value of zero in all the surrounding grains. Each such order parameter representing a grain has a smooth transition from zero to one across its corresponding diffuse GB. Grains with different orientation in the system are represented with different order parameters, represented by the set ${\boldsymbol\phi}=\{\phi_1, \phi_2, \phi_3,.......,\phi_N\}$, where $N$ is the total number of unique grain-orientations in the microstructure considered. For each $i^{\text{th}}$ grain-orientation, the bulk region is represented by the state $\boldsymbol\phi=\{0,0,...,\phi_i=1,..., 0,0 \}$, whereas its grain boundaries are represented by the state, $0<\phi_i<1$. The numerical values of the order parameters are further bounded by the constraint, $\sum_{i=1}^N \phi_i=1$, that is enforced via the construction of a suitable energy density parametrized in terms of the order parameters. The spatio-temporal evolution of the order parameters, and thereby the evolution of the underlying grain morphology, is governed by the classical Allen-Cahn kinetics \cite{allenCahnEquation}: 
\begin{equation}
     \frac{\partial \phi_i}{\partial t}=-M_{\phi}\Big( \frac{\delta L_{\phi}}{\delta \phi_i} \Big), \quad \text{where}  ~i=1,2...N 
     \label{eqn:GG} 
 \end{equation}
 \noindent where  $L_{\phi}$ is the free energy density (Lagrangian) of the system and $M_\phi$ is the GB mobility. Using standard arguments of reversible thermodynamics, the variational derivative of the Lagrangian is identified as the chemical potential driving the evolution, and is given by:
 \begin{equation*} 
 \mu_{\phi}=\frac{\delta L_{\phi}}{\delta \phi}
 \label{eqn:mu}
 \end{equation*}
 The free energy of the system that is minimized by the above kinetics is given by:
 \begin{equation*}
     \Pi=\int_{\Omega} L_{\phi} ~dV=\int_{\Omega} \Big[ g(\boldsymbol\phi) + \sum_{i=1}^{N} \frac{\epsilon}{2} |\nabla\phi_i|^2\;\Big] dV
 \end{equation*} 
 \noindent In the above equation, the first term of the Lagrangian, $g(\boldsymbol\phi)$, is a double-well potential representing the driving force for grain growth, and the second term represents the \textcolor{black}{gradient energy term} parametrized by the interface parameter, $\epsilon$. The specific form of the double-well potential construction considered here is given by ~\cite{fadi2017immiscibleNC}:
 \begin{equation*}
     g(\boldsymbol\phi)=\frac{4}{3} \Big[ 1- 4 \sum_{i=1}^N \phi_i^3 + 3\big( \sum_{i=1}^N \phi_i^2\big)^2 \Big]
 \end{equation*}
  \noindent Substituting above expressions in Equation~\ref{eqn:GG} yields the governing equation to model multi-grain evolution through the evolution of each order parameter,
 \begin{equation}
     \frac{\partial \phi_i}{\partial t}=-M_{\phi_i} \Big( \frac{\partial \,g(\boldsymbol\phi)}{\partial\, \phi_i} -\epsilon \sum_{i=1}^{N} \nabla^2\phi_i \Big), \quad \text{where}  ~i=1,2...N
     \label{eqn:allenCahn}
 \end{equation} 
 
 The above governing equations are a standard treatment for modeling grain growth, and these equations minimize the free energy by minimizing the GB energy through grain coarsening, i.e., by reducing the fraction of GBs compared to the bulk. This model is relevant to grain growth and recrystallization in traditional microcrystalline metallic alloys. However, in nanocrystalline alloys, additional physical mechanisms of GB segregation and solute drag caused by solute precipitation counter this tendency for grain coarsening. Thus, to model microstructure evolution in nanocrystalline alloys, it's important to effectively represent these mechanisms. Further, it is also important to model mechanical deformation in conjunction with these mechanisms to study the physically relevant problem of grain size stabilization under load.  The incorporation of these additional mechanisms along with finite-strain mechanics is treated below.\\

\subsection{Numerical basis of modeling GB segregation and solute drag}
On top of the standard grain growth model described in the previous section, we aim to incorporate the ability to model three phenomena: \textcolor{black} { Solute drag caused by (1) GB segregation of solute and (2) Solute precipitation at GB and junctions, and (3) Finite-strain mechanics for modeling deformation}. As stated earlier, the utility of these three models is in studying the effect of these physical phenomena in stabilizing grain sizes in nanocrystalline alloys, against the natural tendency for grain coarsening in regular metallic alloys. 
  
 \begin{figure}[!h]
    \centering
    \includegraphics[width=1.0\linewidth]{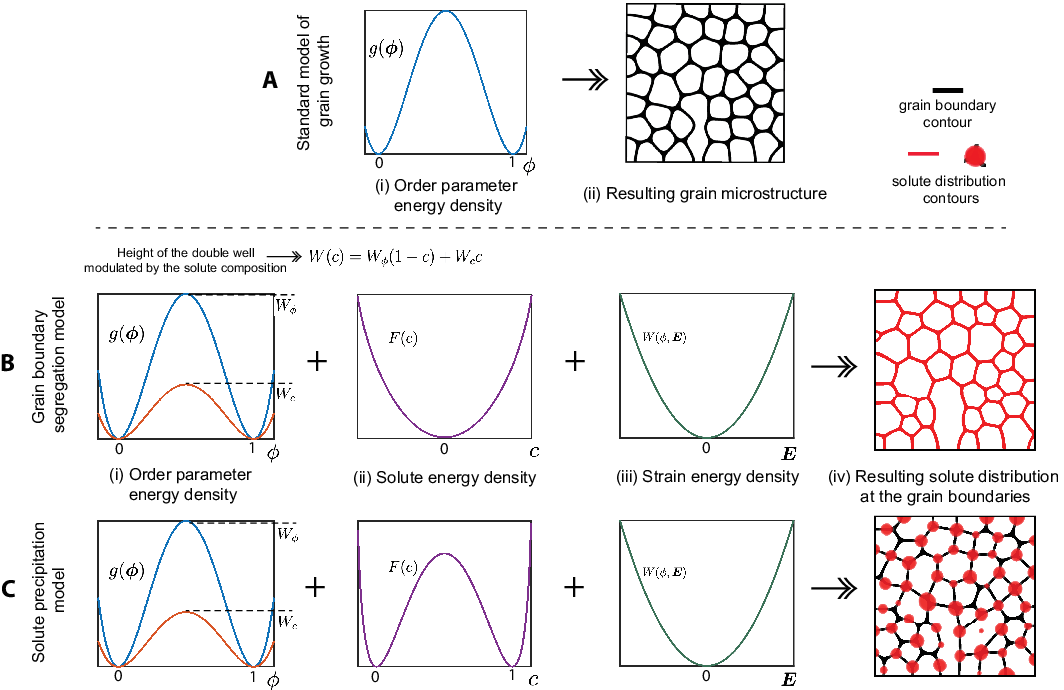}
    \caption{Schematics showing the free energy density construction for modeling (A) Standard grain growth, (B) GB segregation, and (C) Solute precipitation. For the later two cases, representative free energy density constructions for the (i) order parameter, (ii) solute composition, and (iii) mechanics fields are shown in the first, second and third column, respectively. Also shown in the fourth column are the (iv) resulting grain microstructure and solute distribution at the GBs and junctions.}
    \label{fig:schematicFreeEnergy}
\end{figure}

The central tenet of the phase-field model is first order kinetics driven by free energy minimization. In the standard model of grain growth described in the previous section, the system free energy is minimized by material points in the bulk moving to one of the wells, located at $\phi=0$ and $\phi=1$ in the order parameter space, in the double-well characterization of the free energy density.  A schematic showing the construction of these double-well potentials in the order parameter space is shown in Figure \ref{fig:schematicFreeEnergy}. A unified treatment of the construction of these minimizing potentials for the three phenomena of interest listed earlier and its integration with mechanics is a unique aspect of this work. 
   
To stabilize grain growth and avoid further grain coarsening as in nanocrystalline alloys, one needs to make it energetically favorable in the free energy density space for the solute atoms to segregate to the GBs and junctions, and then impede the movement of these GBs. In this section, we will look at two different strategies of modulating the order parameter free energy density as a function of solute composition and also consider the relevant kinetics for the solute atoms. Broadly, this is achieved by modulating a thermodynamic term (double-well height) and a kinetic term (solute mobility) in the coupled phase-field representation of grain structure along with the solute distribution. {\em As stated in the introduction, while some of these strategies have been previously considered in the literature, here we present an integrated framework of both these mechanisms in conjunction with the model for mechanical deformation}.

\paragraph{\bf \small Model of GB segregation} \mbox{}\\
GB segregation is a commonly observed mechanism through which impurity atoms affect the grain growth in a polycrystalline alloy. Impurity atoms via diffusion segregate along the GB and inhibit the GB migration in the system. To develop a thermodynamic model for GB segregation and impurity drag effect, we assume that impurity atoms in the alloy form an ideal solution in which case bulk free energy of the system can be given as,
 \begin{equation}
     G(c,\boldsymbol\phi)=F(c) + g(\boldsymbol\phi) W(c) 
     \label{eqn:GcPhi}
 \end{equation}
here $F(c)$ represents the free energy of the mixing in the alloy which under the ideal solution approximation is given by,
 \begin{equation}
     F(c)=(1-c) G_\phi\, + c\,G_{\text{sol}} + RT \Big( c\, \text{ln}(c) + (1-c)\,\text{ln}(1-c) \Big)
     \label{eqn:FcSegregation}
 \end{equation}
and $W(c)=W_\phi (1-c) + W_c c$ and $g(\boldsymbol\phi)$ represents a solute composition dependent double-well potential. $W(c)$ is constructed so as to energetically prefer lower solute compositions. Such a construction favors solute diffusion towards the GB, and this diffusion is governed by the classical Fick's law,
 
 \begin{equation}
     \frac{\partial c}{\partial t}= - \nabla\cdot \boldsymbol{J}
 \end{equation}
with the flux, {\bf J}, given by,
 \begin{equation}
     \boldsymbol{J} = -c\, (1-c)\, M_{sol}\, \nabla \mu
 \end{equation}
 where $M_{sol}$ is isotropic solute mobility and $\mu$ is the solute chemical potential. With the solution model and the free energy expression considered above, the resulting diffusion equation driving the order parameters is given by,
 \begin{equation}
     \frac{\partial \phi_i}{\partial t}=-M_{\phi_i} \Big(W(c) \frac{\partial g(\boldsymbol\phi)}{\partial \phi_i} -\epsilon \nabla^2\phi_i \Big)
 \end{equation}
 
As the double-well height is now a function of the solute composition, this leads to a dependency of the GB thickness on the solute composition ($\delta=\epsilon/W(c)$). To maintain a fixed GB width we introduce a new scaled parameter $\tilde{\epsilon}=\epsilon/W(c)$. With the introduction of this scaled parameter, the governing equation of grain growth in the presence of a solute distribution is given by,
 \begin{equation}
     \frac{\partial \phi_i}{\partial t}=-M_{\phi_i} W(c) \Big( \frac{\partial g(\boldsymbol\phi)}{\partial \phi_i} - \widetilde{\epsilon}\, \nabla^2 \phi_i \Big)
 \end{equation}
 
 \paragraph{\bf \small Model of solute precipitation} \mbox{}\\
 The second case we consider is a phase separating solution, in which the free energy of the system is given by,
 \begin{equation}
     \Pi=  \int_\Omega L_{c} ~dV = \int_\Omega \Big[ G(c,\boldsymbol\phi) + \frac{\epsilon}{2} |\nabla\phi_i|^2 + \frac{\kappa}{2} |\nabla c|^2 \Big] \; dV
 \end{equation}
 Here $G(c,\boldsymbol\phi)$ has the form given by Equation (\ref{eqn:GcPhi}), but the free energy density of the solution is given by, 
 \begin{equation}
     F(c)= (1-c) G_{\boldsymbol\phi} + cG_{sol} + RT \Big(c\,ln(c) + (1-c)\,ln(1-c) \Big) + \Omega_{\text{mix}}\,c(1-c)
     \label{eqn:regularSolution}
 \end{equation}
 where $\Omega_{mix}c(1-c)$ is the heat of mixing. The evolution of the solute field is governed through the classical Cahn-Hilliard~\cite{cahnHilliard1958} equation,
 \begin{equation} 
     \frac{\partial c}{\partial t}= \nabla\cdot \Big[M_{sol} \nabla \Big(\frac{\delta L_{c}}{\delta c} \Big) \Big]
 \end{equation}
This equation leads to fourth order spatial derivatives on the composition variable $c$, so to make this more suitable for the numerical formulation considered later in this work, we choose to instead write this equation as two separate coupled equations,
 \begin{equation}
     \frac{\partial c}{\partial t}=M_{sol}\nabla^2\mu
     \label{eqn:CH1}
 \end{equation}
 \begin{equation}
     \mu=\frac{\partial G(c,\phi)}{\partial c} - \kappa\nabla^2 c
          \label{eqn:CH2}
 \end{equation}

 \paragraph{\bf \small Model of mechanical deformation} \mbox{}\\
Considering the framework of finite-strain mechanics, total free energy of the system with the addition of the elastic strain energy density is given by,
\begin{equation}
    \Pi= \int_\Omega L_{\bf u}~dV= \int_\Omega\Big[ \textcolor{black}{g}(\boldsymbol\phi) + \frac{\epsilon}{2} \sum_{i=1}^N |\nabla\phi_i|^2 + \frac{1}{2}\, \boldsymbol E: \mathbb{C(\boldsymbol\phi)}:\normalsize{\boldsymbol{E}} \Big] \; dV
\end{equation}
where $\boldsymbol E$ is the Green-Lagrange strain metric and $\mathbb{C(\boldsymbol\phi)}$ is an average elasticity tensor that is expressed a function of the elasticity tensors corresponding to each grain orientation, $\mathbb{C}_i$, and is given by,
\begin{equation}
    \mathbb{C(\boldsymbol\phi)}= \frac{\sum_{i=1}^N h(\phi_i) \mathbb{C}_i } {\sum_{i=1}^N h(\phi_i) }
\end{equation}
where $h(\phi_i)={\big(1+ sin(\pi (\phi_i- \frac{1}{2}))\big)}/2$ is a smooth function of grain orientations \cite{tonks2011polycrystalline}. 

Ignoring inertial effects and body forces, the governing equation of mechanical equilibrium is given by,
\begin{equation}
    \boldsymbol\nabla\cdot \boldsymbol P =0 
\end{equation}
where ${\boldsymbol P} = \frac{\delta L_{\bf u}}{\delta \boldsymbol F } $ is the first Piola-Kirchhoff stress that here is a function of the strain as well as the underlying grain orientations.
\par
\textcolor{black}{
Previously we looked at three phenomenon namely GB segregation, solute precipitation and mechanical deformations. However, in a realistic system these phenomenon occur simultaneously. From modeling perspective GB segregation and solute precipitation are similar phenomenon difference being in the interaction between solute particles and base metal. As discussed previously when solute particles and base metal form an ideal solution we observe GB segregation, while solute precipitation is observed when a regular solution is formed. Henceforth, we present a coupled model which incorporated the previously discussed phenomenon.}

 \paragraph{\bf \small \textcolor{black}{Coupled model of grain growth with GB segregation and mechanical deformation}} \mbox{}\\
 \textcolor{black}{Assuming solute particles form an ideal solution with the base metal and system is mechanically deformed, total free energy of the system, considering finite strain mechanics, can be given as,
\begin{equation}
	\Pi=\int_{\Omega}L\; dV= \int_{\Omega} \Big[ G(c, \boldsymbol\phi) + \frac{\epsilon}{2} \sum_{i=1}^{N} |\nabla\phi_i|^2 +\frac{1}{2} \boldsymbol E : \mathbb{C}(\boldsymbol\phi):\boldsymbol E \Big]\; dV
	\label{eqn:freeEnergyGBSegregationAndMechanics}
\end{equation}
here $G(c, \boldsymbol\phi)$ due to assumption of ideal solution model takes the same form as equation (\ref{eqn:GcPhi}) and (\ref{eqn:FcSegregation}), and can be given in a complete form as,
\begin{equation}
	G(c, \boldsymbol\phi)=(1-c)G_{\phi} + c G_{\text{sol}} + RT \Big(c\,\text{ln}(c) + (1-c)\, \text{ln}(1-c)\Big) + g(\boldsymbol\phi) W(c)
\end{equation}
Following the same assumptions regarding mechanical deformations and same treatment of interfacial energy parameter as presented previously, governing equations for evolutions of order parameter and solute, and governing equation of mechanical equilibrium can be given as,
\begin{equation}
\begin{aligned}
	\frac{\partial \phi_i}{\partial t}&= -M_{\phi_i} W(c)  \Big( \frac{\partial g(\boldsymbol\phi)}{\partial \phi_i} -\tilde\epsilon \nabla^2\phi_i \Big) -\frac{M_{\phi_i}}{2}\Big( \boldsymbol E: \frac{\partial \mathbb{C(\boldsymbol\phi)}}{\partial \phi_i}:\boldsymbol E \Big)\\
	\frac{\partial c}{\partial t}&= c(1-c)M_{\text{sol}} \nabla^2\mu \\
	\mu&= \frac{\partial F(c)}{\partial c} + g(\boldsymbol\phi) \frac{\partial W(c)}{\partial c} \\
	&\nabla\cdot \boldsymbol P=0
\end{aligned}
\label{eqn:strongFormMechanicsSegregation}
\end{equation}
}
 
 \paragraph{\bf \small \textcolor{black}{Coupled model of grain growth with solute precipitation and mechanical deformation}} \mbox{}\\
 \textcolor{black}{ With the assumption of a regular solution model, total free energy of the mechanically deformed system, considering finite strain mechanics, can be given as,
 \begin{equation}
 \Pi=\int_{\Omega} L\;dV=\int_{\Omega} \Big[  G(c, \boldsymbol\phi) + \frac{\epsilon}{2} \sum_{i=1}^{N} |\nabla\phi_i|^2 +\frac{\kappa}{2} |\nabla c|^2+\frac{1}{2} \boldsymbol E : \mathbb{C}(\boldsymbol\phi):\boldsymbol E  \Big]\;dV
 \label{eqn:freeEnergyPrecipitationAndMechanics}
 \end{equation} 
 here $G(c, \boldsymbol\phi)$ under the assumption of regular solution model takes the form as presented in equation (\ref{eqn:GcPhi}) and (\ref{eqn:regularSolution}), and can be given as,
 \begin{equation}
 G(c, \boldsymbol\phi) = (1-c) G_{\boldsymbol\phi} + cG_{\text{sol}} + RT \Big(c\,\text{ln}(c) + (1-c)\,\text{ln}(1-c) \Big) + \Omega_{\text{mix}}\,c(1-c) + g(\boldsymbol\phi)W(c)
 \end{equation}
 Based on the given free energy expression governing equations for evolution for order parameter and solute evolution, and governing equations for mechanics can be given as,
 \begin{equation}
 \begin{aligned}
 	\frac{\partial \phi_i}{\partial t}&= -M_{\phi_i} W(c)  \Big( \frac{\partial g(\boldsymbol\phi)}{\partial \phi_i} -\tilde\epsilon \nabla^2\phi_i \Big) -\frac{M_{\phi_i}}{2}\Big( \boldsymbol E: \frac{\partial \mathbb{C(\boldsymbol\phi)}}{\partial \phi_i}:\boldsymbol E \Big) \\
	\frac{\partial c}{\partial t}&= M_{\text{sol}} \nabla^2\mu \\
	\mu&= \frac{\partial G(c, \boldsymbol\phi)}{\partial c} -\kappa \nabla^2 c \\
	&\nabla\cdot \boldsymbol P=0
 \end{aligned}
 \label{eqn:strongFormMechanicsPrecipitation}
 \end{equation}
 }

\section{Variational Formulation}
We now pose the above governing equations in their weak (integral) form. This formulation is used to solve these equations within a standard finite element framework. \\ 
Find the primal fields $\{ \phi_i, c, \mu, \bf u \}$,
\begin{align*}
 \phi_i &\in \mathscr{S}_{\phi_i},  \quad \mathscr{S}_{\phi_i} = \{  \phi_i \in \text{H}^1(\Omega) ~\vert  ~ \phi_i  = ~ \phi_i' ~\forall ~\textbf{X} \in \Gamma^{ \phi_i} \}, \\
c &\in \mathscr{S}_{c },  \quad \mathscr{S}_{c } = \{ c  \in \text{H}^1(\Omega) ~\vert  ~c  = ~c' ~\forall ~\textbf{X} \in \Gamma^{c} \}, \\
\mu &\in \mathscr{S}_{\mu},  \quad \mathscr{S}_{\mu} = \{ \mu \in \text{H}^1(\Omega) ~\vert  ~\mu  = ~\mu' ~\forall ~\textbf{X} \in \Gamma^{\mu} \}, \\
{\bf u} &\in \mathscr{S}_{\bf u},  \quad \mathscr{S}_{\bf u} = \{ {\bf u} \in \text{H}^1(\Omega) ~\vert  ~{\bf u} = ~{\bf u}' ~\forall ~\textbf{X} \in \Gamma^{\bf u} \} 
\end{align*}
such that for their respective variations $\{ w, \tilde{w}, \nu, \bf w \}$,   
\begin{align*}
\forall ~ w &\in \mathscr{V}_{\phi_i},  \quad \mathscr{V}_{\phi_i} = \{  w \in \text{H}^1(\Omega) ~\vert  ~ w  = ~ 0 ~\forall ~\textbf{X} \in \Gamma^{ \phi_i} \}, \\
\forall ~ \tilde{w} &\in \mathscr{V}_{c },  \quad \mathscr{V}_{c } = \{ \tilde{w}   \in \text{H}^1(\Omega) ~\vert  ~\tilde{w}   = ~0 ~\forall ~\textbf{X} \in \Gamma^{c} \}, \\
\forall ~\nu &\in \mathscr{V}_{\mu},  \quad \mathscr{V}_{\nu} = \{ \nu \in \text{H}^1(\Omega) ~\vert  ~\nu  = ~0 ~\forall ~\textbf{X} \in \Gamma^{\mu} \}, \\
\forall ~{\bf w} &\in \mathscr{V}_{\bf u},  \quad \mathscr{V}_{\bf u} = \{ {\bf w} \in \text{H}^1(\Omega) ~\vert  ~{\bf w} = ~0 ~\forall ~\textbf{X} \in \Gamma^{\bf u} \} 
\end{align*}
we satisfy the following integral equations in each of the different cases listed below,

\paragraph{Grain boundary segregation}
Weak form of the equations for the order parameter and solute:
 \begin{equation}
 \begin{aligned}
     \int_\Omega w \frac{\partial \phi_i}{\partial t}\; dV&= -M_{\phi_i} W(c) \int_{\Omega} w\, \frac{\partial g(\boldsymbol\phi)}{\partial \phi_i}\;dV - M_{\phi_i}W(c)\,\widetilde{\epsilon}\int_\Omega \nabla w\cdot \nabla\phi\;dV\\
     &\int_{\Omega} \tilde{w} \frac{\partial c}{\partial t}\, dV=-M_{sol} \int_{\Omega}c\,(1-c)\, \nabla\tilde{w}\cdot \nabla\mu\; dV\\
     &\int_{\Omega} \nu\, \mu\, dV=\int_{\Omega} \nu \frac{\partial F(c)}{\partial c}\; dV + \int_{\Omega} \nu \,g(\phi) \frac{\partial W(c)}{\partial c}\; dV
    \end{aligned}
    \label{eqn:weakFormSegregation}
 \end{equation}
 
 \paragraph{Solute precipitation}
 Modified weak form of the equations for only the solute: 
 \begin{equation}
 \begin{aligned}
     \int_\Omega \tilde{w} \frac{\partial c}{\partial t}\; dV= -M_{sol}\int_\Omega \nabla\tilde{w}\cdot\nabla\mu\; dV \\
     \int_\Omega \nu\mu\;dV=\int_\Omega \nu \frac{\partial G(c,\boldsymbol\phi)}{\partial c}\; dV +\kappa \int_\Omega \nabla\nu\cdot\nabla c\; dV
 \end{aligned}
     \label{eqn:weakFormPrecipitation}
 \end{equation}
 
 \paragraph{Mechanics driven grain growth}
Weak form of the equations for order parameter evolution with mechanical equilibrium:
 \begin{equation}
 \begin{aligned}
     \int_\Omega w \frac{\partial \phi_i}{\partial t} dV = -M_{\phi_i} \int_\Omega w \frac{\partial g(\boldsymbol\phi)}{\partial\phi_i} dV -& M_{\phi_i}\epsilon\int_\Omega \nabla w \cdot \nabla\phi_i dV - \frac{M_{\phi_i}}{2}\int_\Omega w \Big(\boldsymbol E:\frac{\partial \mathbb{C(\boldsymbol\phi)}}{\partial \phi_i}:\boldsymbol E \Big) dV \\
     &\int_\Omega \nabla{\bf w}\cdot{\bf P} dV=0
 \end{aligned}
     \label{eqn:weakFormMechanics}
 \end{equation}
 
\textcolor{black}{ 
  \paragraph{Mechanics driven grain growth with GB segregation}
  Weak form of the equation(\ref{eqn:strongFormMechanicsSegregation}) for evolution of order parameter, solute, and mechanical equilibrium:
  \begin{equation}
  \begin{aligned}
  	 \int_\Omega w \frac{\partial \phi_i}{\partial t} dV = -M_{\phi_i}W(c) \int_\Omega& w \frac{\partial g(\boldsymbol\phi)}{\partial\phi_i} dV-M_{\phi_i}W(c)\tilde\epsilon\int_\Omega \nabla w \cdot \nabla\phi_i dV-\frac{M_{\phi_i}}{2}\int_\Omega w \Big(\boldsymbol E:\frac{\partial \mathbb{C(\boldsymbol\phi)}}{\partial \phi_i}:\boldsymbol E \Big) dV \\
	 &\int_{\Omega} \tilde{w} \frac{\partial c}{\partial t}\, dV=-M_{sol} \int_{\Omega}c\,(1-c)\, \nabla\tilde{w}\cdot \nabla\mu\; dV\\
	 &\int_{\Omega} \nu\, \mu\, dV=\int_{\Omega} \nu \frac{\partial F(c)}{\partial c}\; dV + \int_{\Omega} \nu \,g(\phi) \frac{\partial W(c)}{\partial c}\; dV \\
	 &\int_\Omega \nabla{\bf w}\cdot{\bf P} dV=0
  \end{aligned}
  \end{equation}
  }
  
  \textcolor{black}{
  \paragraph{Mechanics driven grain growth with solute precipitation}  
  Weak form of the equation(\ref{eqn:strongFormMechanicsPrecipitation}) for evolution of order parameter, solute, and mechanical equilibrium:
  \begin{equation}
  \begin{aligned}
  	 \int_\Omega w \frac{\partial \phi_i}{\partial t} dV = -M_{\phi_i}W(c) \int_\Omega& w \frac{\partial g(\boldsymbol\phi)}{\partial\phi_i} dV - M_{\phi_i}W(c)\tilde\epsilon\int_\Omega \nabla w \cdot \nabla\phi_i dV - \frac{M_{\phi_i}}{2}\int_\Omega w \Big(\boldsymbol E:\frac{\partial \mathbb{C(\boldsymbol\phi)}}{\partial \phi_i}:\boldsymbol E \Big) dV \\
	 & \int_\Omega \tilde{w} \frac{\partial c}{\partial t}\; dV= -M_{sol}\int_\Omega \nabla\tilde{w}\cdot\nabla\mu\; dV \\
	  &\int_\Omega \nu\mu\;dV=\int_\Omega \nu \frac{\partial G(c,\boldsymbol\phi)}{\partial c}\; dV +\kappa \int_\Omega \nabla\nu\cdot\nabla c\; dV \\
	 &\int_\Omega \nabla{\bf w}\cdot{\bf P} dV=0
  \end{aligned}
  \end{equation}
 In all cases, no flux boundary conditions are considered for the order parameters and solute fields.  }
 
 \section{Computational implementation}
Time integration of diffusion equations uses the backward Euler method. In the following sections $\Delta t=t^{n+1}-t^n$ will be used to denote the time step. Here, we present a time-discrete form of Equation (\ref{eqn:weakFormSegregation}) only, however similar forms can be obtained for Equation (\ref{eqn:weakFormPrecipitation}) and Equation (\ref{eqn:weakFormMechanics})  as well,
 \begin{equation}
     \begin{aligned}
     &\int_{\Omega} w (\phi_i^{n+1}-\phi_i^n)\;dV= -M_\phi W(c^{n+1})\Delta t \int_\Omega \Big(\frac{\partial g(\boldsymbol\phi^{n+1})}{\partial \phi_i^{n+1}} + \widetilde{\epsilon}\; \nabla w\cdot \nabla\phi_i^{n+1} \Big)dV\\
     &\int_{\Omega} \tilde{w} (c^{n+1}-c^n)\, dV=-M_{sol}\Delta t \int_{\Omega}c^{n+1}\,(1-c^{n+1})\, \nabla\tilde{w}\cdot \nabla\mu^{n+1}\; dV\\
     &\int_{\Omega} \nu\, \mu^{n+1}\, dV=\int_{\Omega} \nu F'(c^{n+1})\; dV + \int_{\Omega} \nu \,g(\boldsymbol\phi^{n+1}) W'(c^{n+1})\; dV
     \end{aligned}
 \end{equation}
 A $C^0$-continuous basis code based on the standard Finite Element Method (FEM) is used to solve the above weak forms of the governing equations. The code base is an in-house, C++ programing language based, parallel code framework with adaptive meshing and adaptive time-stepping -  build on top of the deal.II open source Finite Element library \cite{dealII95}.
 
\section{Results}
We now present simulations of the numerical formulations of grain growth, solute segregation and drag, and mechanical deformation. In each case, we model grain growth and investigate how GB evolution or its stabilization is achieved - with and without external loads. While the formulations presented in this work are for three dimensional problem geometries, considering the computational complexity of these simulations with numerous order parameters, we initially focus on 2D grain evolution examples, but later also discuss a 3D example (albeit a thin 3D slice of material). The polycrystalline grain structures considered have about 60 grains with six different sets of grain orientations that are randomly assigned to the grains. When present, the solute atoms are initially considered as homogeneously distributed over the entire domain. For the 2D simulations, the computational discretization consider a mesh of 128$\times$128 elements, and in 3D a similar mesh density is considered with a few elements along the thickness direction of the thin material slices. To quantify the evolution of the GBs, we use the GB energy as a measure of the total GB length in a domain at any instance. This is a good measure of the total grain length as the GB energy is directly proportional to the GB length under the assumption of a fixed GB thickness. \\

\subsection{Grain Boundary Segregation} \label{GBsegregation}

\begin{figure}[!h]
    \centering
    \subfloat[]{
        \includegraphics[scale=0.45]{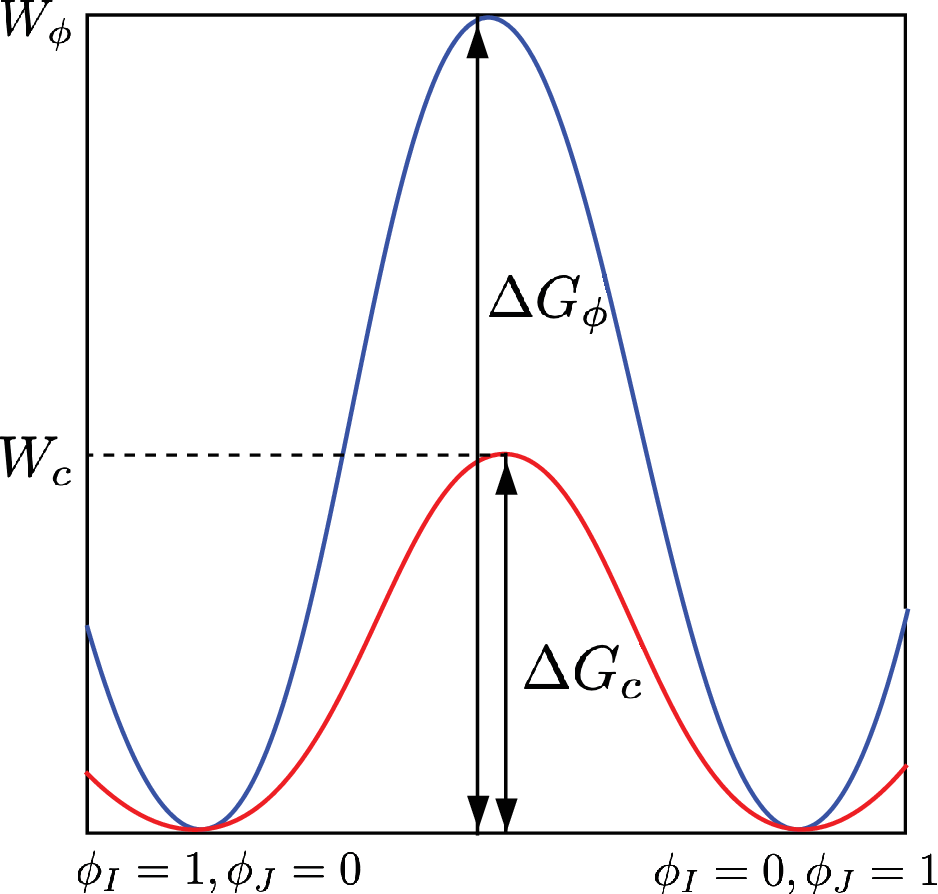}
        \label{fig:drivingForce}
    }
    \quad
    \subfloat[]{
        \includegraphics[scale=0.45]{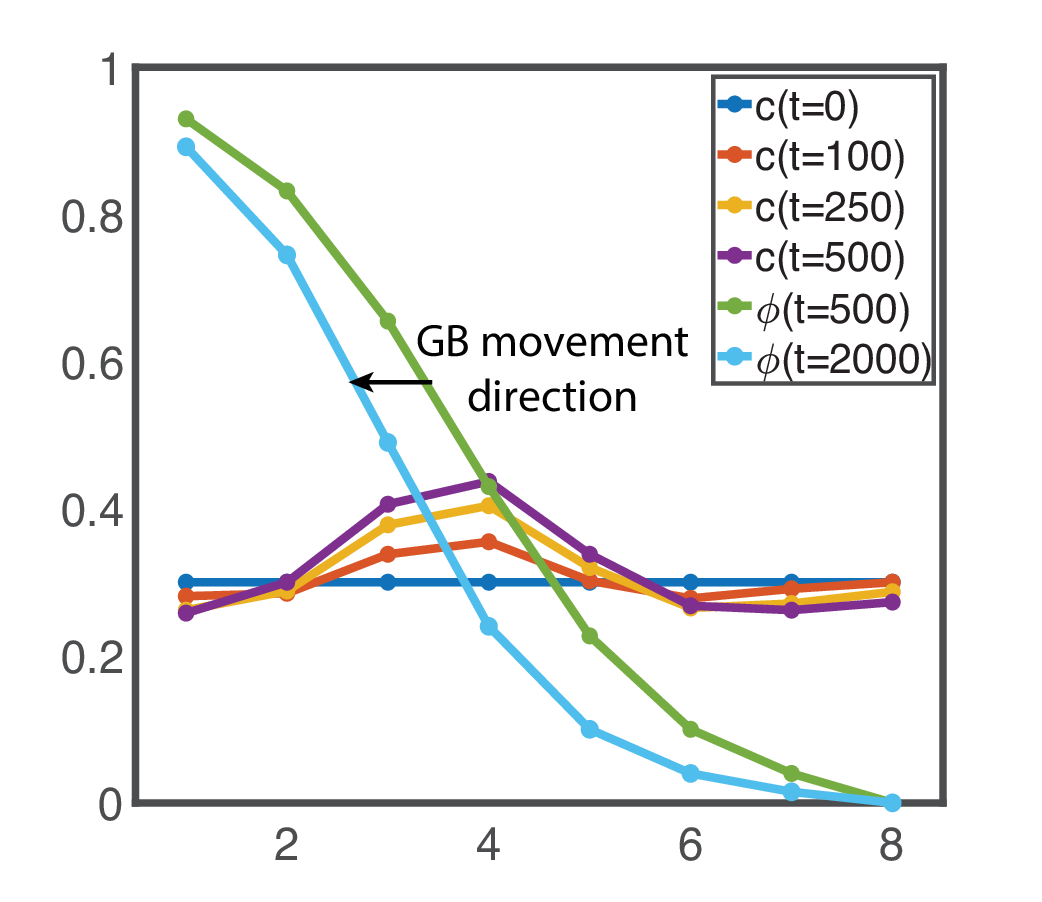}
        \label{fig:soluteProfile}
    }
    \caption{(a) Double-well potential and driving force for grain growth without interaction of solute atoms (blue) and with GB segregation (red), (b) \textcolor{black}{Solute composition across GB at different time instances and asymmetry of the solute profile across GB caused by GB movement.}}
    \label{fig:my_label}
\end{figure}

We now focus on the thermodynamics and kinetics of GB segregation and the drag effect created on GB migration by solute particles. Figures \ref{fig:drivingForce} and Figures \ref{fig:soluteProfile} are instructive for the discussion presented. In the formulation presented, GB segregation is achieved via the solute composition dependent double-well potential. Double-well potential, $g({\boldsymbol{\phi}})$, by construction results in a positive contribution to the free energy in the GB region (implicitly delineated using the gradient of the order parameters), and a zero contribution in bulk region. The height of the double-well potential depends upon solute composition, $W(c)$. Ideally, $W(c)$ would be estimated using molecular dynamics simulation or from experimental observations as different impurity/solute atoms may induce a different grain growth driving force. In this work, we have chosen a linear dependence,  $W(c)=W_\phi(1-c) + cW_c $, and it has been set up as a decreasing function of solute composition by the choice of the parameters $W_\phi=1$ and $W_c=-0.1$. This drives the solute atoms to diffuse into the GB region via Fickian diffusion to attain a lower free energy for the system. Here, $W_\phi$ and $W_c$ are the heights of double-well potential in case of solute free GB and solute rich GB, respectively. These also represent the driving force for grain growth in the system($\Delta G_\phi$ and $\Delta G_c$ in Figure \ref{fig:drivingForce}). Thus, solute segregation into GB creates a drag effect which reduces the overall driving force for grain growth. The kinetics of the drag effect can be better understood by examining solute composition across the GB. \textcolor{black}{Figure \ref{fig:soluteProfile} shows solute composition at different nodes across the GB in a FE simulation, and it can be seen that, with time, solute particles diffuse from the bulk region to the GB region and segregate along the GB. As a result solute composition increases in the GB and reduces in the bulk region and the solute forms a symmetric composition profile across a stationary GB($\phi(t=500)$). However, during GB migration($\phi(t=2000)$), solute composition profile across GB becomes asymmetric which causes a drag effect on the moving GB, reducing GB migration speed. As the GB migrates, it drags the segregated solute with it, and if solute mobility is comparably lower than the GB mobility the asymmetry becomes prominent. This effect might not be evident, if solute mobility is kept to be of the same order as GB mobility in the simulation. In summary, in the presented model free energy has been setup in a way that effect of solute composition on double-well potential directly reflects on the GB mobility, and thus the slow kinetics of the GB is achieved through the thermodynamics. }
\begin{figure}[!h]
    \centering
    \includegraphics[scale=0.5]{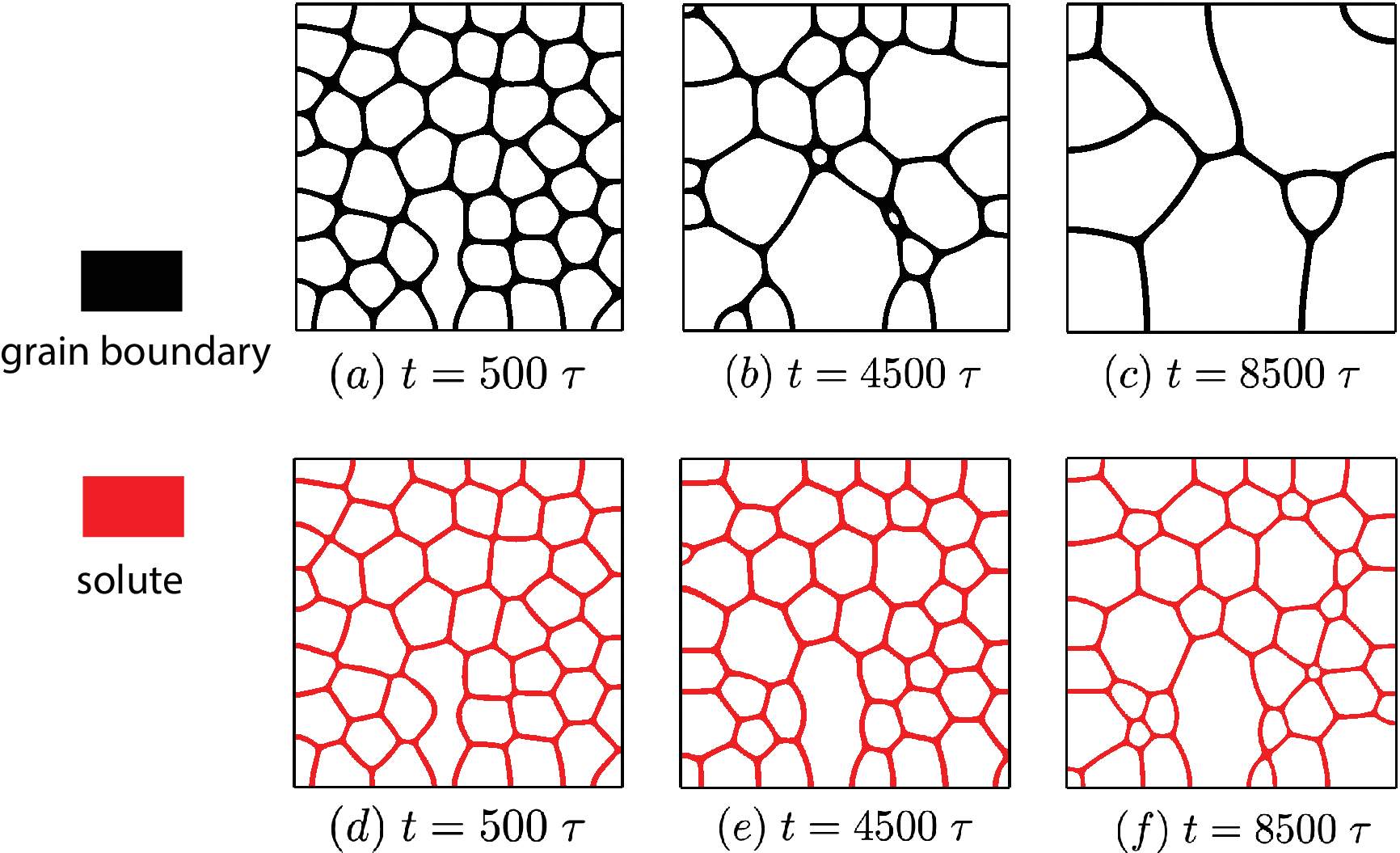}
    \caption{Comparison of grain structure evolution at three different time steps without interaction of the solute atoms (a,b,c) and with the drag effect due to GB segregation (d,e,f).}
    \label{fig:segregation}
\end{figure}

\par
\begin{figure}[!h]
    \centering
    \includegraphics[scale=0.7]{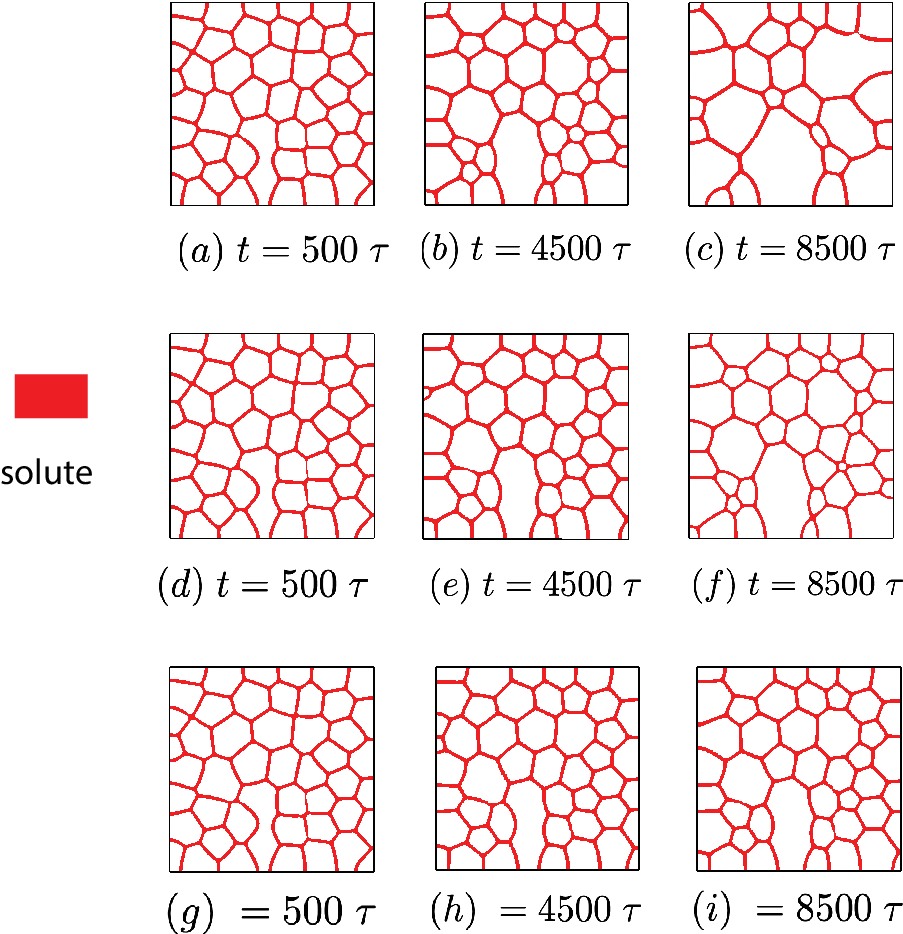}
    \caption{Comparison of evolving grain structure with different drag force exerted on GB, controlled by $W_c$; $W_c=0.2$ (a,b,c), $W_c=-0.1$ (d,e,f), $W_c=-0.4$(g,h,i)}
    \label{fig:parametric}
\end{figure}

Figure \ref{fig:segregation} shows a comparison of evolving grain structure in the absence and the presence of GB segregation in a polycrystalline NC alloy. Here, we assume initial state of solute to be homogeneously distributed in the system with a composition of 0.3. Then we allow the solute to diffuse for about 500 time steps in the equilibration stage of the simulation during which solute atoms segregate to the GB. Figure~\ref{fig:segregation}(d,e,f) show the GBs after solute has segregated into the GB. It should be clear from the discussion presented above that $W(c)$ plays an important role in modeling GB segregation and the solute drag effect. To understand its effect in more detail, we investigate grain growth in NC alloys with different values of $W_c$. Figure~\ref{fig:parametric} shows evolution of grain structure over time for different values of $W_c$ ($W_c=0.2, -0.1$ and $-0.4$). We see that $W_c=-0.4$ offers the maximum drag effect as the thermodynamic driving force for grain growth decreases the most in this case.

Further, to quantify the solute drag effect, we compute total GB energy ($\frac{\epsilon}{2} \sum_{i=1}^N|\nabla\phi|^2$) in the system. Figure \ref{fig:segregationPlot} shows a comparison of GB energy in the system with and without GB segregation. It can be seen that GB energy decreases much slowly in case of GB segregation as compared to grain growth without any segregation effect. The initial increase in the GB energy in case of segregation is during the equilibration time for GB stabilization. Similarly Figure~\ref{fig:parameterPlot} shows GB energy variation with time for different values of $W_c$, and it can be seen that $W_c=-0.4$ offers maximum drag effect as GB energy variation is slowest with time as compared to $W_c=0.2$ and $W_c=-0.1$.

\subsection{Solute Precipitation}\label{precipitation}
In the previous section, we studied the effects of GB segregation which assumes an ideal solution model. However, NC alloys mostly form a regular solution in which the system phase separates into solute rich and solute depleted zones through spinodal decomposition. 
\begin{figure}[!h]
\hspace{-0.2in}
    \centering
    \includegraphics[scale=0.5]{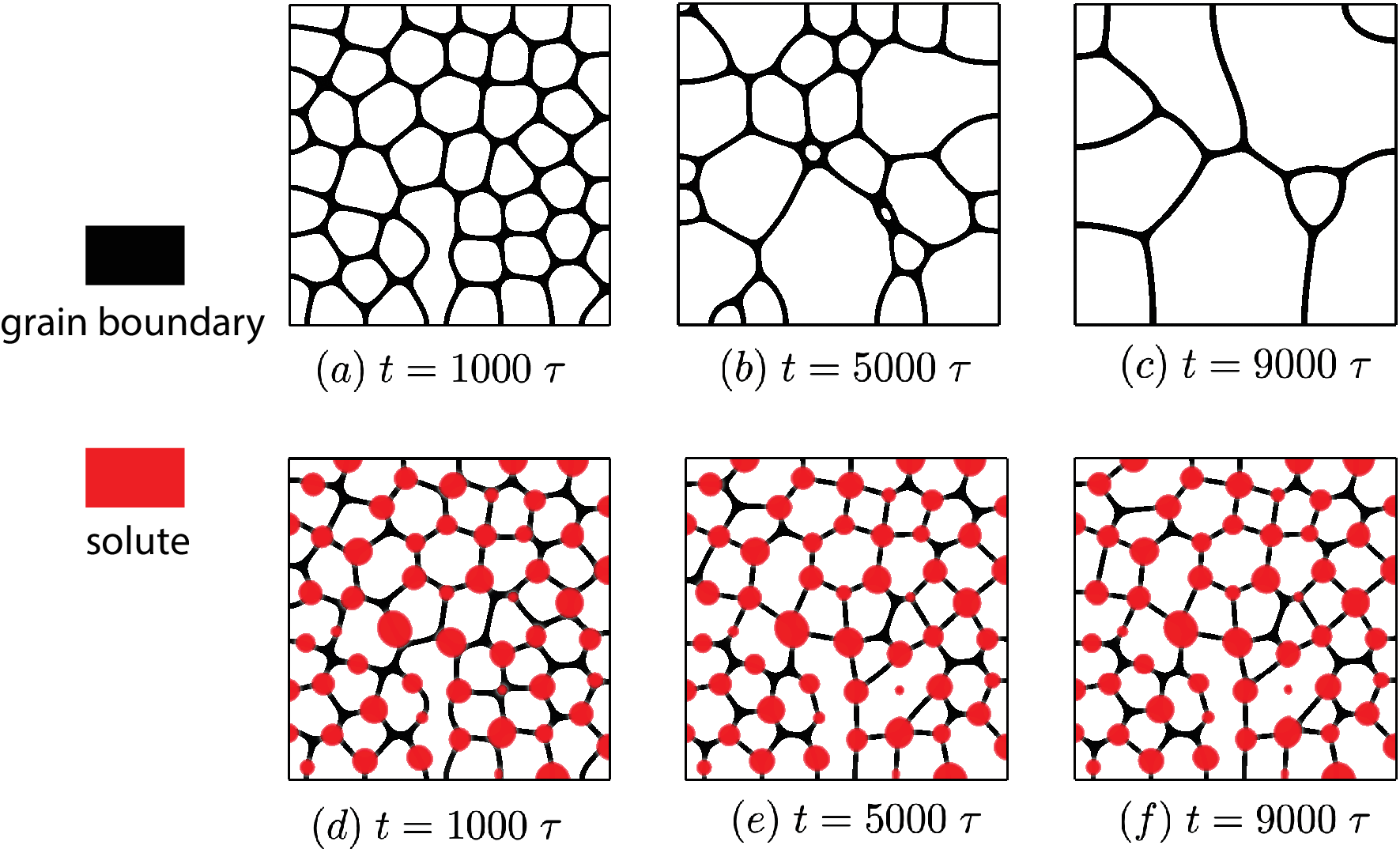}
    \caption{Comparison of grain structure at different time steps without soute interaction (a,b,c) and with solute precipitaiton( d,e,f), demonstrating the pinning behavior of the solute precipitates at the triple junctions.}
    \label{fig:precipitation}
\end{figure}

\par 
Figure~\ref{fig:precipitation} presents a comparison of evolving grain structure with time in the presence and absence of solute precipitation. Similar to the case of GB segregation, we allow for the solute to evolve in the system through Equations \ref{eqn:CH1} and \ref{eqn:CH2}. We observe solute precipitating at the triple junctions of the GBs, as the double-well potential, $g({\boldsymbol{\phi}})$, by construction attains its maxima at the triple junctions. To minimize the total free energy of the system, solute atoms start to diffuse towards the GBs and the triple junctions, and eventually form precipitates at the triple junctions. As shown in Figures~\ref{fig:precipitation}(d,e,f), the solute atoms initially precipitate to the GBs. However, with the passage of time, the solute atoms eventually diffuse towards the triple junctions to form precipitates. 

\par
The thermodynamic arguments about the drag force presented in the previous results section for GB segregation remain valid for solute precipitate as well. However, here the drag effect is mostly localized to the GB triple junctions to which the solute has precipitated. This is unlike the case of GB segregation discussed earlier where the drag force acts on the entire GB. As the solute atoms precipitate to the triple junctions, they impose a pinning effect on the GBs and prevent the triple junctions to migrate. This eventually leads to the fixation of all the triple junctions in the grain structure, and the solute-free GBs can only evolve between the fixed triple junctions. This leads to the stabilization of the NC alloy grain structure. The temporal evolution of the GB energy also supports this argument. As can be seen in Figure~\ref{fig:precipitationPlot}, for the case of solute precipitation, the GB energy stagnates in the system after some time - indicating a stable grain structure.

\subsection{Effect of mechanical deformation}\label{deformation}

\begin{figure}[!h]
\centering
\includegraphics[scale=0.25]{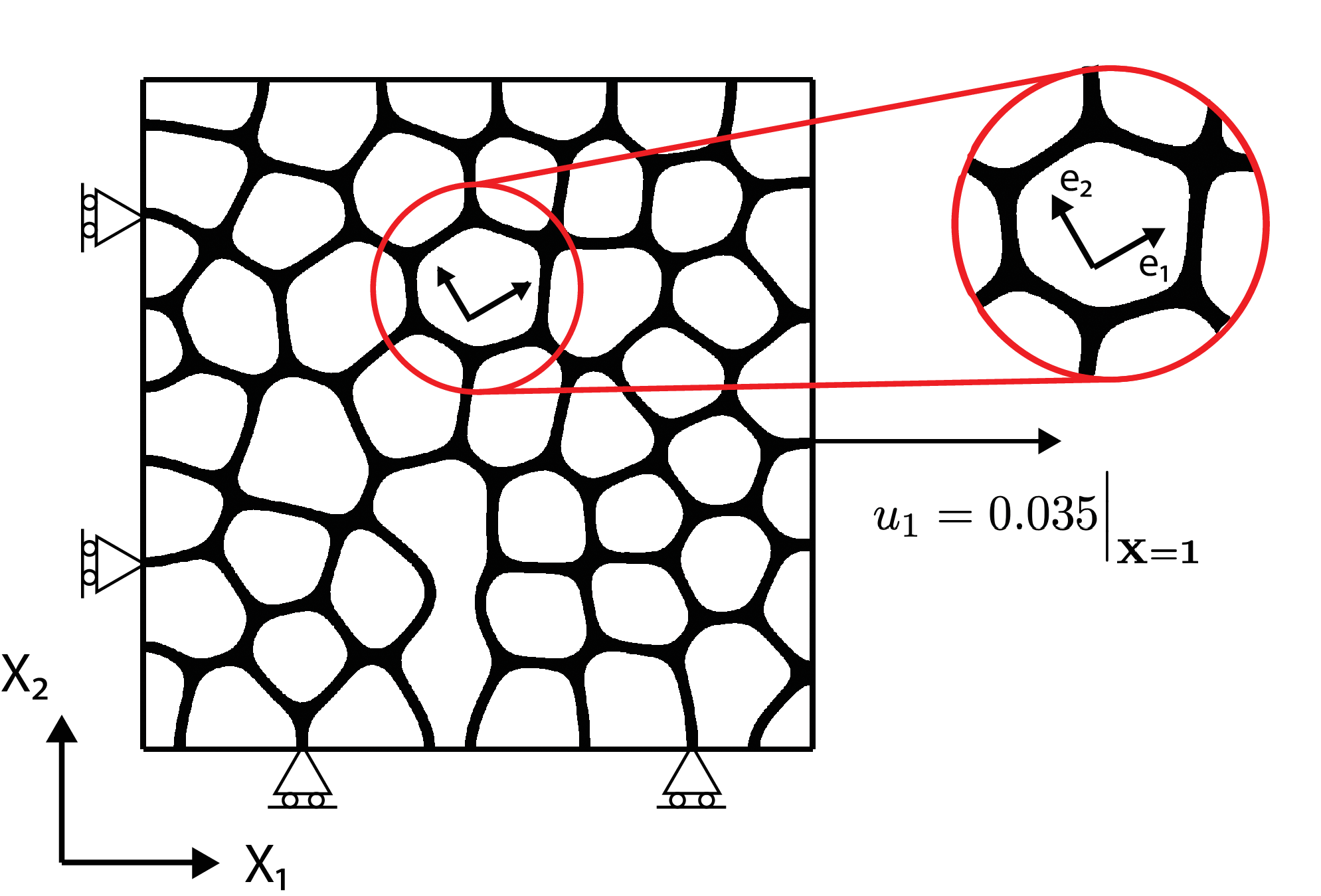}
\caption{Displacement boundary conditions applied on the geometry. Also shown in the inset image are the local crystallographic direction of the grains.}
\label{fig:boundaryCondition}
\end{figure}

In previous sections, we looked at the drag effect caused by impurity/solute atoms present in the system. In this section, we examine the effects of external loading on grain growth. For this case, we apply a Dirichlet boundary condition on the displacement at the right end, as shown in Figure~\ref{fig:boundaryCondition}. We consider an elastic anisotropy for the grains such that the elastic modulus in the ${\bf e_2}$ direction is twice its value along the ${\bf e_1}$ direction, where ${\bf e_1}$ and ${\bf e_2}$ are the local crystallographic orientation of each grain. Due to elastic anisotropy considered, in addition to GB energy minimization, grain growth is now also driven so as to minimize the stored elastic strain energy. This results in differential deformation of the grains along the different crystallographic axes, and this imposes an additional driving force on the GBs. As will be shown, we observe that the rate of grain growth increases due to this additional driving force from strain energy minimization under an external load.

\begin{figure}[!h]
    \centering
    \includegraphics[scale=0.5]{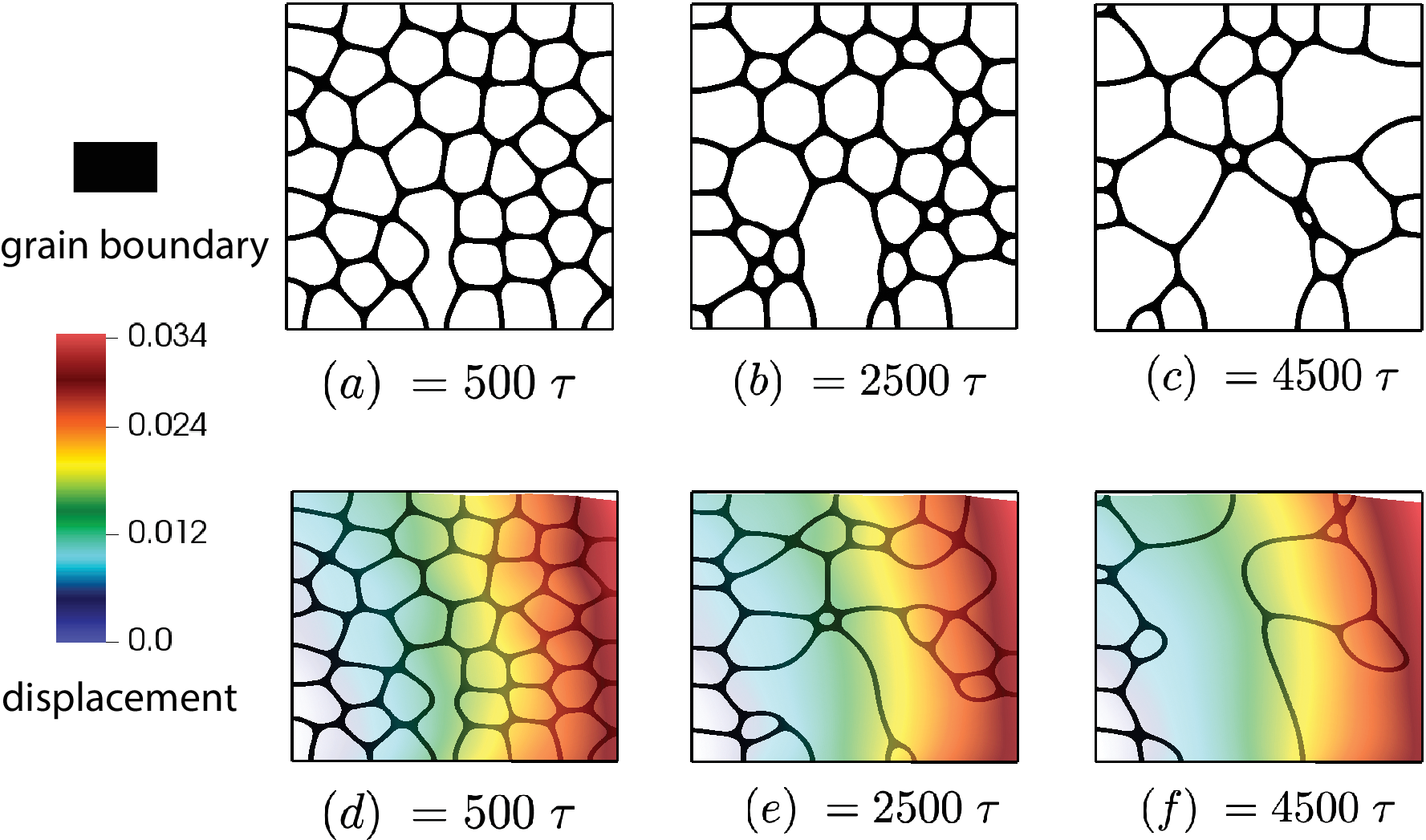}
    \caption{Comparison of grain structure without load (a,b,c), and with the application of uniaxial tensile load (d,e,f).}
    \label{fig:mechanics}
\end{figure}

Figure~\ref{fig:mechanics} presents a comparison of evolving grain structure with and without the effects of external load. To observe the effects of external load on grain growth, we apply desired boundary conditions before the GBs are numerically permitted to migrate (non-zero GB mobility). 

\par
Under influence of an applied load, grain growth is driven by minimization of the GB energy as well as minimization of the elastic strain energy. In this context, $\int_{\Omega}\big( \frac{\partial G(\boldsymbol{\phi})}{\partial \phi_i}-\epsilon \nabla^2\phi_i\big)dV$ denotes the chemical energy contribution for grain growth, and $\frac{1}{2}\int_{\Omega}\big({\bf E}:\frac{\partial \mathbb{C}(\boldsymbol{\phi})}{\partial \phi_i}:{\bf E}\big)dV$ denotes strain energy contribution for grain growth. Depending on their individual contributions, GB energy and strain energy may either complement each other or compete against each other. If $\int_{\Omega}\Big(\frac{\partial G(\phi)}{\partial \phi_i}-\epsilon\nabla^2\phi_i\Big) \;dV > \frac{1}{2}\int_{\Omega}{\bf E}:\frac{\partial \mathbb{C}(\phi)}{\partial\phi_i}:{\bf E}\;dV$, strain energy has a significantly lower contribution towards grain growth than GB energy, and the underlying microstructure of the NC alloy remains only minimally affected. However, if $\int_{\Omega}\Big( \frac{\partial G(\phi)}{\partial \phi_i}-\epsilon\nabla^2\phi\Big) \;dV < \frac{1}{2}\int_{\Omega}{\bf E}:\frac{\partial \mathbb{C}(\phi)}{\partial\phi_i}:{\bf E}\;dV$, grain growth is  driven by mechanics and we see a grain structure for which stored elastic strain energy decreases more rapidly than the GB energy, resulting in a higher rate of grain growth. 
This effect can be seen by examining the grain structure in Figure~\ref{fig:mechanics}, as well as by comparing the total GB energy of the system in Figure~\ref{fig:mechanicsPlot}. As has been observed in the simulations, during grain growth, grains with crystal orientations that favor minimum strain energy are preferred - resulting in a directionally aligned microstructure. As a consequence, since mechanical properties of NC alloys are a result of the underlying grain microstructure, preferred grain orientations (texture) can be achieved though mechanical loading during fabrication of NC alloy parts \textcolor{black}{\cite{tonks2011polycrystalline, tonksMillet2010}}.
 
\subsection{Solute drag effect in the presence of deformation}
\label{dragMechanics}
In the previous sections, we have separately considered the effects of GB segregation, solute precipitation and mechanics driven grain growth. However, in the physical system, all these effects play out together and are coupled. Thus, to better appreciate grain growth and grain stabilization in NC alloys, it is important to look at the combined effects of these phenomena. Here, we will consider a 3D geometry with both the effects of solute interaction and mechanics on GB evolution. As mentioned earlier, the formulations presented in this work are for 3D problem domains, however, considering the computational complexity of the simulations with numerous order parameters, we initially focussed on 2D grain evolution examples, but now present a 3D example to demonstrate geometric generality of this framework. We will consider two different cases: (I) GB segregation with an applied load, and (II) Solute precipitation with an applied load, both on a 3D domain with the same initial grain structure as the 2D examples considered in the previous sections.

\begin{figure}[!h]
    \centering
    \includegraphics[scale=0.25]{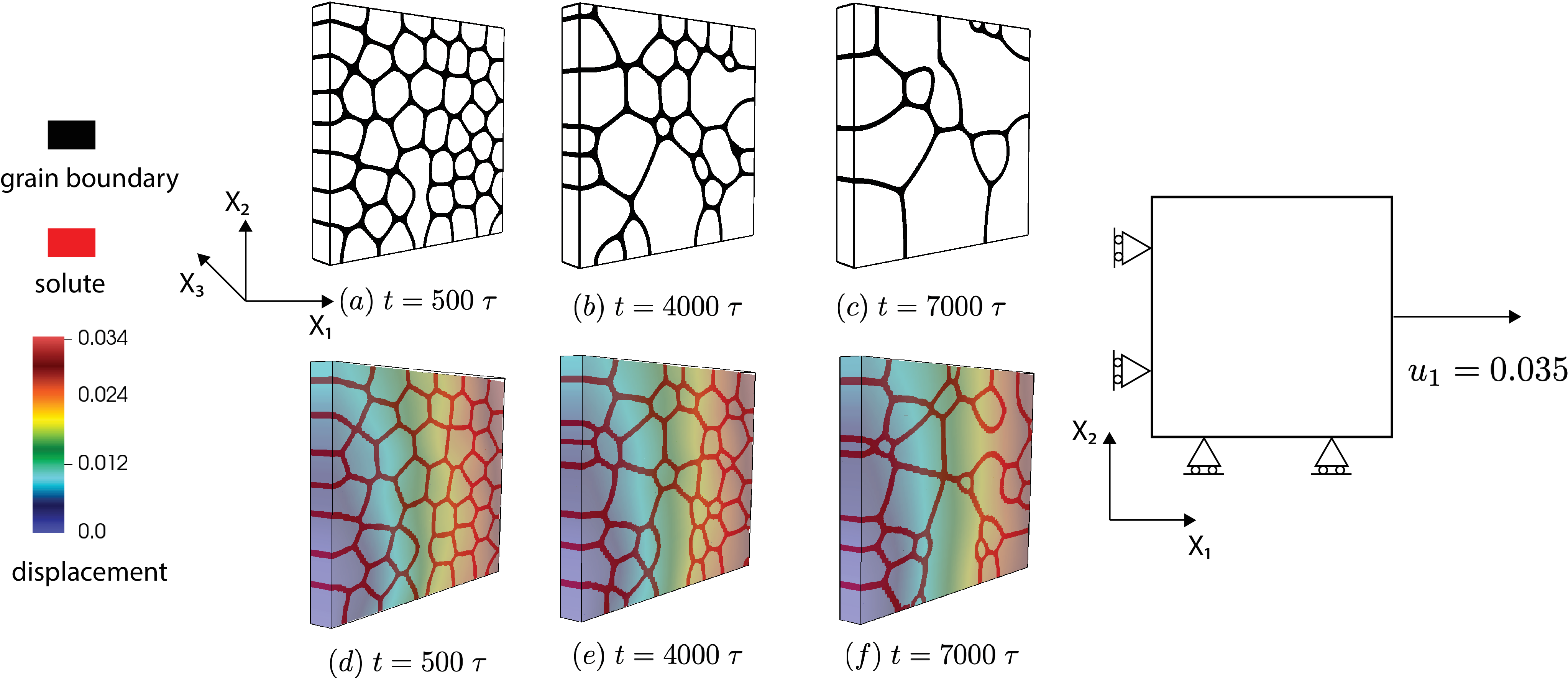}
    \caption{Grain growth with the solute drag effect due to GB segregation and mechanics (d,e,f), and its comparison with standard grain growth without GB segregation and mechanics (a,b,c). }
    \label{fig:segregationMechanics}
\end{figure}

\par
Figure~\ref{fig:segregationMechanics} shows a comparison of evolving grain structure in Case (I) (GB segregation with mechanics) - grain growth without solute precipitation. Here, GB segregation and mechanics are competing against each other. Grain growth is encouraged due to strain energy minimization, but the drag force on the moving grain boundary from GB solute segregation impedes faster grain growth as discussed in Section~\ref{deformation}. However, elastic strain energy does drive the direction in which GBs migrate. By examining the GB energy variation in the system (Figure~\ref{fig:segregationMechanicsPlot}) we can observe that the rate of grain growth has indeed been reduced, but the solute drag effect is not as strong as seen in Section~\ref{GBsegregation} due to the additional driving force provided by strain energy minimization.

\begin{figure}[!h]
    \centering
    \includegraphics[scale=0.25]{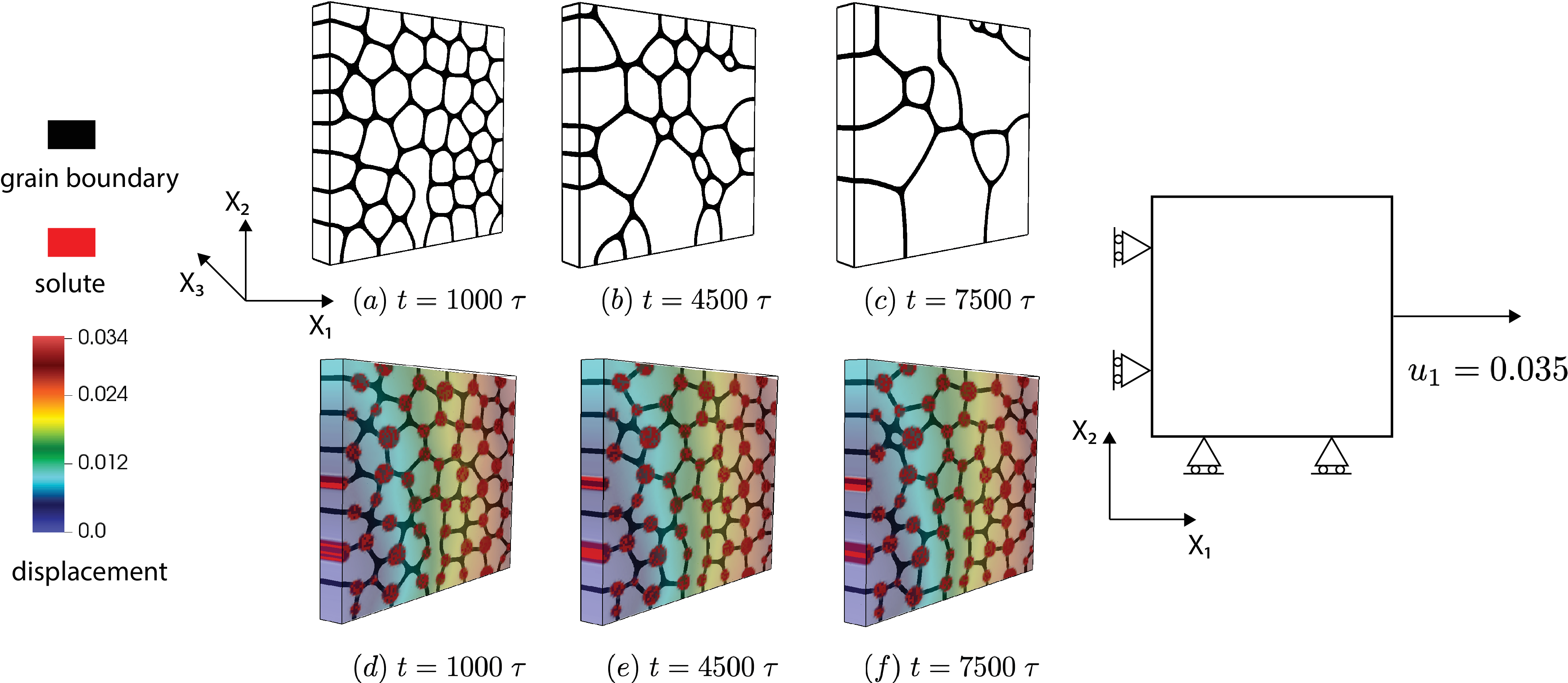}
    \caption{Grain growth with the solute drag effect due to solute precipitation and mechanics (d,e,f), and its comparison with standard grain growth without solute precipitation and mechanics (a,b,c).}
    \label{fig:precipitationMechanics}
\end{figure}

\par
Figure~\ref{fig:precipitationMechanics} presents grain growth simulation for Case (II) (Solute precipitation with mechanics) - grain growth with solute precipitation. We observe that as in Section~\ref{precipitation}, solute atoms precipitate to the triple junctions and have a pinning effect on these junctions. Although an external load has been applied on the system, stored strain energy does not have much effect on grain growth as the triple junctions become immobile due to solute pinning effect - resulting in a very stable grain structure. This fact can be further appreciated by observing the GB energy variation in the system (Figure~\ref{fig:precipitationMechanicsPlot}) which shows that the GBs have fully stabilized, and thus, the GB energy remains constant.

\begin{textblock}{20}(12,6.5)

\end{textblock}

\pgfplotstableread{2D_segregation_withoutDrag_energy.txt}{\PhaseFieldSeg}
\pgfplotstableread{2D_segregation_withDrag_energy.txt}{\segregation}

\pgfplotstableread{Wc_comparison_02_energy.txt}{\WCa} 
\pgfplotstableread{Wc_comparison_01_energy.txt}{\WCb} 
\pgfplotstableread{Wc_comparison_04_energy.txt}{\WCc} 

\pgfplotstableread{2D_precipitation_withoutDrag_energy.txt}{\PhaseFieldPrec}

\pgfplotstableread{2D_precipitation_withDrag_energy.txt}{\cahnHilliard}

\pgfplotstableread{2D_mechanics_withoutMechanics_energy.txt}{\PhaseFieldMech}

\pgfplotstableread{2D_mechanics_finiteStrain_energy.txt}{\mechanicsFinite}

\pgfplotstableread{2D_mechanics_smallStrain_energy.txt}{\mechanicsSmall}

\pgfplotstableread{3D_withoutSegregationMechanics_energy.txt}{\PhaseFieldSegMech}

\pgfplotstableread{3D_segregationMechanics_energy.txt}{\mechSeg}

\pgfplotstableread{3D_withoutPrecipitationMechanics_energy.txt}{\PhaseFieldPrecMech}

\pgfplotstableread{3D_precipitationMechanics_energy.txt}{\mechPrec}

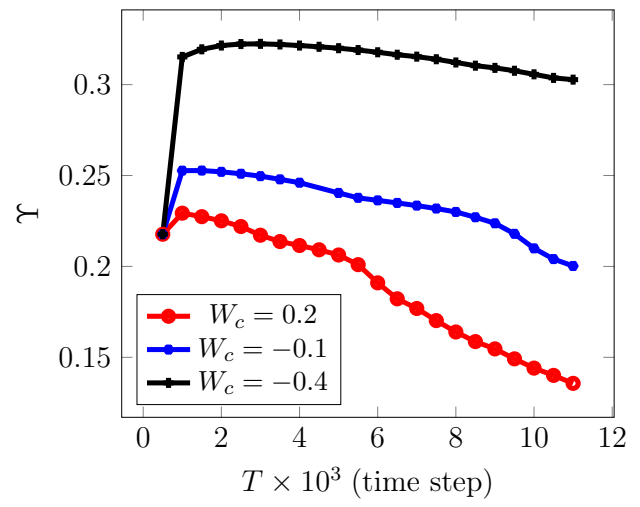
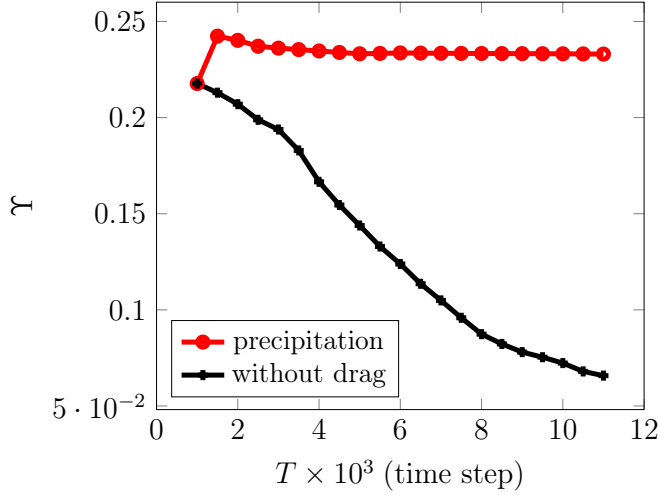
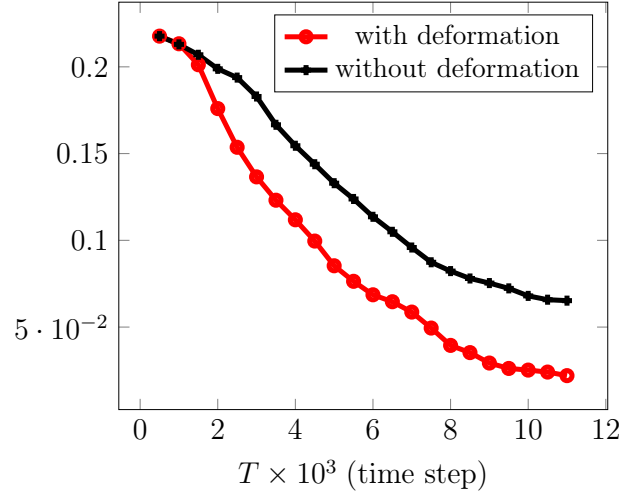
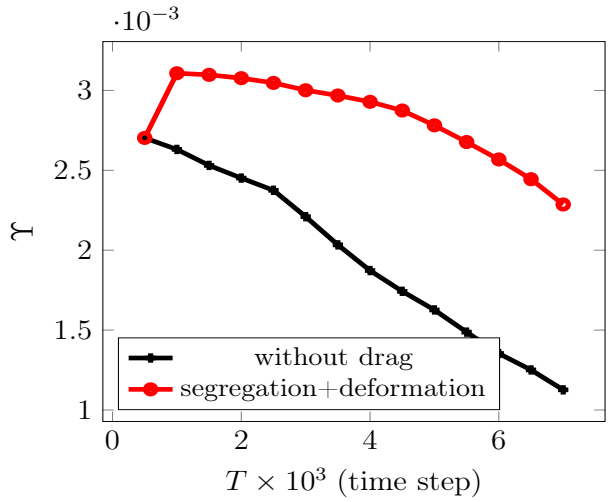
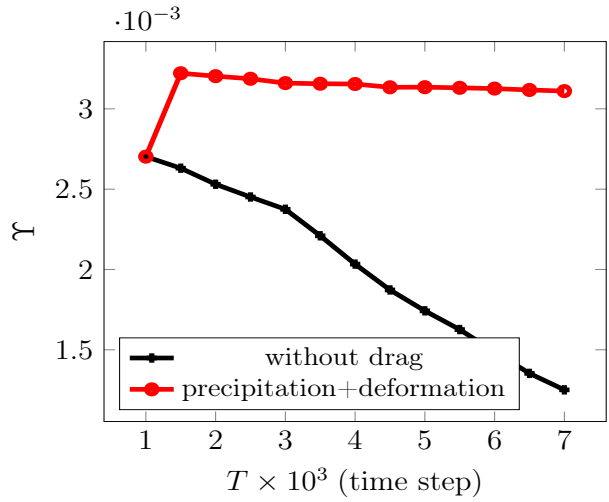
\begin{figure}[!ht]
\centering
\vspace{-0.5in}
\subfloat[]{
\hspace{-1.0in}
\resizebox{0.55\textwidth}{0.4\textwidth}{
    \begin{tikzpicture}[scale=1]
    \begin{axis}[xlabel={$T\times 10^3\;$(time step)},ylabel={$\Upsilon$}, legend pos=south west]
    \addplot [red, mark=o,line width=2.0pt] table [x={var1}, y={var4}] {\segregation};
    \addlegendentry{\small{GB segregation}}
    \addplot [black, mark=+,line width=2.0pt] table [x={var1}, y={var4}] {\PhaseFieldSeg};
    \addlegendentry{\small{without drag}}
    \end{axis}
  
    \end{tikzpicture}
    }
    \label{fig:segregationPlot}
}
\quad
\subfloat[]{
\hspace{-0.1in}
\resizebox{0.52\textwidth}{0.4\textwidth}{
    \begin{tikzpicture}[scale=1]
    \begin{axis}[xlabel={$T\times 10^3\;$(time step)},ylabel={$\Upsilon$}, legend pos=south west]
    \addplot [red, mark=o,line width=2.0pt] table [x={var1}, y={var4}] {\WCa};
    \addlegendentry{\small{$W_c=0.2$}}
    \addplot [blue, mark=asterisk,line width=2.0pt] table [x={var1}, y={var4}] {\WCb};
    \addlegendentry{\small{$W_c=-0.1$}}
    \addplot [black, mark=+,line width=2.0pt] table [x={var1}, y={var4}] {\WCc};
    \addlegendentry{\small{$W_c=-0.4$}}
    \end{axis}

    \end{tikzpicture}
    }
    \label{fig:parameterPlot}
}

\vspace{0.0 in}
\subfloat[]{
\hspace{-1.0in}
\resizebox{0.55\textwidth}{0.4\textwidth}{
\begin{tikzpicture}[scale=1]
    \begin{axis}[xlabel={$T\times 10^3\;$(time step)},ylabel={$\Upsilon$},legend pos=south west]
    \addplot [red, mark=o,line width=2.0pt] table [x={var1}, y={var4}] {\cahnHilliard};
    \addlegendentry{\small{precipitation}}
    \addplot [black, mark=+,line width=2.0pt] table [x={var1}, y={var4}] {\PhaseFieldPrec};
    \addlegendentry{\small{without drag}}
    \end{axis}
  
    \end{tikzpicture}
    }
    \label{fig:precipitationPlot}
}
\quad
\subfloat[]{
\hspace{-0.5in}
\resizebox{0.55\textwidth}{0.4\textwidth}{
\begin{tikzpicture}[scale=1]
    \begin{axis}[xlabel={$T\times 10^3\;$(time step)},ylabel={$\Upsilon$}, legend pos=north east]
    \addplot [red, mark=o,line width=2.0pt] table [x={var1}, y={var4}] {\mechanicsFinite};
    \addlegendentry{\small{with deformation}}
    \addplot [black, mark=+,line width=2.0pt] table [x={var1}, y={var4}] {\PhaseFieldMech};
    \addlegendentry{\small{without deformation}}
    \end{axis}

    \end{tikzpicture}
    }
    \label{fig:mechanicsPlot}
}

\vspace{0.1in}
\subfloat[]{
\hspace{-0.8in}
\resizebox{0.5\textwidth}{0.4\textwidth}{
\begin{tikzpicture}[scale=1]
    \begin{axis}[xlabel={$T\times 10^3\;$(time step)},ylabel={$\Upsilon$},legend pos=south west]
    \addplot [black, mark=+,line width=2.0pt] table [x={var1}, y={var4}] {\PhaseFieldSegMech};
    \addlegendentry{\small{without drag}}
    \addplot [red, mark=o,line width=2.0pt] table [x={var1}, y={var4}] {\mechSeg};
    \addlegendentry{segregation+deformation}
    
    \end{axis}
   
    \end{tikzpicture} 
    }
    \label{fig:segregationMechanicsPlot}
}
\quad
\subfloat[]{
\resizebox{0.5\textwidth}{0.4\textwidth}{
\begin{tikzpicture}[scale=1]
    \begin{axis}[xlabel={$T\times 10^3\;$(time step)},ylabel={$\Upsilon$},legend pos=south west]

    \addplot [black, mark=+,line width=2.0pt] table [x={var1}, y={var4}] {\PhaseFieldPrecMech};
    \addlegendentry{\small{without drag}}
    \addplot [red, mark=o,line width=2.0pt] table [x={var1}, y={var4}] {\mechPrec}; 
    \addlegendentry{precipitation+deformation}
    
    \end{axis}
  
    \end{tikzpicture}
    }
    \label{fig:precipitationMechanicsPlot}
}
\caption{GB energy variation in the system for (a) GB segregation in 2D (Section~\ref{GBsegregation}), (b) for different values of $W_c$ in GB segregation in 2D (Section~\ref{GBsegregation}), (c) solute precipitation in 2D (Section~\ref{precipitation}), (d) applied load in 2D (Section~\ref{deformation}), (e) GB segregation with applied load in 3D (Section~\ref{dragMechanics}), and (f) solute precipitation with applied load in 3D (Section~\ref{dragMechanics})}
\label{fig:energyPlots}
\end{figure}

 \clearpage
\section{Conclusion}
\label{conclusion}
In this work, we have modeled important phenomena that are known to affect grain growth in nanocrystalline alloys. To this end, we have considered the effects of GB solute segregation, solute precipitation, and mechanical deformation. We have developed a coupled phase-field model, considering a solute composition dependent double-well potentials, and studied their effect on grain growth. We have considered both an ideal solution model to represent solute behavior in the dilute limit to model GB segregation, and a regular solution model to explain solute precipitation when there exists a miscibility gap and system undergoes a phase separation. We carried out a parametric studies to estimate the effect of the solute drag force on the GB movement in cases of GB segregation and solute precipitation. We also present a mechano-chemical phase-field model to understand the effects of mechanical deformation on grain growth in NC alloys. We observe that GB segregation and solute precipitation offer a drag force on the moving GB, in turn reducing the grain growth driving force, thus, stabilizing the grain structure in NC alloys. However, under the effect of an applied load, the rate of grain growth increases due to the additional driving force on GBs from strain energy minimization. For a broader understanding of microstructure stabilization, we also present studies of the coupled effects of GB segregation, solute precipitation and mechanical loading. This work numerically demonstrates that in NC alloys, grain size stabilization can be achieved through impurity/solute atoms, and that the orientations of the stabilized grain microstructures can be modulated by applying mechanical deformation.

\bibliographystyle{elsarticle-num-names}
\bibliography{ main.bib}
\end{document}